\documentstyle[epsf]{kluwer}
\def\gapprox{\;\rlap{\lower 2.5pt            
 \hbox{$\sim$}}\raise 1.5pt\hbox{$>$}\;}       
\def\lapprox{\;\rlap{\lower 2.5pt            
 \hbox{$\sim$}}\raise 1.5pt\hbox{$<$}\;} 
\def\N{\,{\rm I\kern-.20em N}}
\begin{document}

\begin{article}
\begin{opening}
\title{Are there Radio-quiet Solar Flares?}
\author{Arnold O. \surname{Benz} $^1$ }
\author{Roman \surname{Braj\v sa}$^2$}
\author{Jasmina \surname{Magdaleni\'c}$^2$}
\institute{$^1$ Institute of Astronomy, ETH Z\"urich, CH-8092 Z\"urich, Switzerland (e-mail: benz at astro.phys.ethz.ch)\\
$^2$ Hvar Observatory, Faculty of Geodesy, University of Zagreb, Ka\v ci\'ceva 26, HR-10000 Zagreb, Croatia\\}
\runningtitle{Radio-quiet Solar Flares}
\runningauthor{Arnold O. Benz, Roman Braj\v sa and Jasmina Magdaleni\'c}
\date{Received: 30 November 2006; accepted: 8 January 2007}
\begin{abstract} Some 15\% of solar flares having a soft X-ray flux above GOES class C5 are reported to lack coherent radio emission in the 100 -- 4000 MHz range (type I -- V and decimetric emissions). A detailed study of 29 such events reveals that 22 (76\%) of them occurred at a radial distance of more than 800$''$ from the disk center, indicating that radio waves from the limb may be completely absorbed in some flares. The remaining seven events have statistically significant trends to be weak in GOES class and to have a softer non-thermal X-ray spectrum. All of the non-limb flares that were radio-quiet $>$ 100 MHz were accompanied by metric type III emission below 100 MHz. Out of 201 hard X-ray flares, there was no flare except near the limb ($R>800''$) without coherent radio emission in the entire meter and decimeter range. We suggest that flares above GOES class C5 generally emit coherent radio waves when observed radially above the source. 

\end{abstract} 
\end{opening}
\section{Introduction}
Solar flares release a considerable fraction of their energy into electron acceleration (Brown, 1971; Strong {\it et al.}, 1984; Emslie {\it et al.}, 2005; Saint-Hilaire and Benz, 2005). High-energy electrons propagate until they collide with ambient particles in the corona or reach a layer of higher density, emitting bremsstrahlung in hard X-rays (HXR). Thus flares are generally associated with enhanced hard X-ray emission, although such emission may not always be observable with present sensitivity. The HXR fluence is a good proxy for the energy in accelerated electrons and is taken here as the reference for the time line of the impulsive phase and flare energy release.

Historically however, the flare radio emission was the first signature of non-thermal electrons (Wild, 1950; Boischot and Denisse, 1957). Thus a correlation between radio flux and X-ray emission was sought, as soon as the latter became available. Early attempts were summarized by Kundu (1965), and the relation was reviewed later by Trottet (1986), and Bastian, Benz, and Gary (1998). It soon became clear that a detailed correlation exists between the hard X-rays and the centimeter radio emission, generally attributed to gyrosynchrotron emission (Takakura, 1959). Kosugi,  Dennis, and Kai (1988) derived a linear relation between peak fluxes in hard X-rays ($>$ 30 keV) and gyrosynchrotron emission at 17 GHz.  Extended events (duration longer than ten minutes) have stronger centimeter emission by a factor of two -- ten (Kai, Kosugi, and Nitta, 1985). Although the gyrosynchrotron emission originates from mildly relativistic electrons and hard X-rays were measured at lower energies, one may expect that every flare emits radio emission above the electron gyrofrequency, starting at around 1 GHz and peaking between 3 and beyond 10 GHz. Gyrosynchrotron emission has indeed been reported from the smallest flares known, occurring in quiet regions (Krucker and Benz, 2000). Therefore, it is likely that all hard X-ray emitting flares emit gyrosynchrotron emission in centimeter waves at some level.

Here, the focus is on different radio emission processes: coherent radio emission at meter and decimeter wavelengths ({\it e.g.} McLean and Labrum, 1985). It has a much narrower bandwidth and is often strongly polarized. Coherent emission can reach enormous flux levels during flares, occasionally exceeding a thousand times the quiet-Sun level at decimeter wavelength and a hundred thousand times at meter wavelengths. The great efficiency of coherent emission is the result of the combined action of many electrons organized by kinetic plasma waves. Such waves are believed to be driven by a non-maxwellian distribution of the electron velocity, such as beams, electric currents or trapped non-thermal particles (textbooks by Melrose, 1980; Benz, 2002). Small coherent radio bursts at the limit of present routine observations emit an energy of some $10^{15}$erg. Even if the conversion of particle energy into radio emission occurs at a rate of only $10^{-6}$ (Lin {\it et al.}, 1986; Robinson, Cairns, and Willes, 1994), the energy in the electron population that excites coherent radio emission is a tiny fraction of the flare energy. Thus a slight deviation from an isotropic maxwellian distribution in the course of acceleration or as a result of particle propagation or magnetic trapping may be sufficient to cause coherent radio emission. 

For this reason, it was to the general surprise that no coherent radio emission was found in 15\% of the events having $>$ 1000 cts s$^{-1}$ at $>$ 25 keV as observed by the Hard X-Ray Burst Spectrometer (HXRBS) on the Solar Maximum Mission (SMM) (Simnett and Benz, 1986). The detection ratio did not improve in more sensitive radio observations: In 34 flares (16.9$\pm$2.9\%) out of the 201 selected X-ray events observed by the Ramaty High-Energy Solar Spectroscopic Imager (RHESSI), no coherent radio emission was associated in Phoenix-2 observations between 100 and 4000 MHz (Benz {\it et al.}, 2005). The existence of such radio-quiet flares puts severe constraints on the acceleration and containment of energetic flare electrons. The enhanced sensitivity and the larger search range of frequencies must have increased the association rate. On the other hand, the extension to smaller events in X-ray magnitude from HXRBS to RHESSI may have decreased the association rate. The two opposing effects have apparently cancelled out. Nevertheless, the authors reported that the radio-quiet events were generally weaker in soft X-rays than the radio-associated ones. The association rate decreases with smaller GOES class.

Here we ask the question: What are the circumstances for radio-quiet flares? The 34 radio-quiet flares of Benz {\it et al.} (2005) are studied here individually and in a detail that was not possible in the large survey.

\section{Observations and Methods}
RHESSI observes the Sun in the energy range from 3 keV to 17 MeV (Lin {\it et al.}, 2002). Imaging is achieved by nine absorbing grids modulating the observed flux by satellite rotation. We use standard RHESSI software (Schwartz {\it et al.}, 2002) and calibrations updated to mid-March 2004. 

The Phoenix-2 spectrometer in Bleien (Switzerland) records continuously the solar radio emission from sunrise to sunset. The full Sun flux density and circular polarization are measured by the frequency-agile receiver in the range from 100 -- 4000 MHz at an integration time of 500 $\mu$s per measurement. Detailed descriptions of the instrument can be found in Benz {\it et al.} (1991) and Messmer, Benz, and Monstein (1999). Since the launch of RHESSI, Phoenix-2 observes an overview from 100 MHz to 4000 MHz in 200 channels known to be low in terrestrial interference. The spectrum was sampled evenly with about an equal number of channels per 1000 MHz. The channel bandwidth varies between one, three, and ten MHz depending on frequency, and the sampling time of each channel was fixed at 100 ms throughout the survey. The bandwidth and background result in a nearly constant sensitivity of about 15 sfu (5 $\sigma$ threshold) in all channels. In the period studied here, the instrument has operated automatically without major interruptions and changes.

Benz {\it et al.} (2005) have selected the flares from the RHESSI flare list in the HESSI Experimental Data Center (HEDC\footnote{HEDC is operated by the Institute of Astronomy at ETH Zurich and can be accessed via the URL {\tt http://www.hedc.ethz.ch/}}, Saint-Hilaire {\it et al.}, 2002) in the period from February 13, 2002, until June 30, 2003. The selection criteria were: {\sl (i)} flares larger than GOES class C5.0 and {\sl (ii)} separated more than ten minutes between peak times. Events within ten minutes were considered as one flare. A total of 670 RHESSI flares were found to satisfy these constraints. {\sl (iii)} From this set, the flares were selected during which Phoenix-2 was operating and observing the Sun. This second step reduced the total of the selected flares for direct comparison to 201. {\sl (iv)} Of all 201 joint flares, Benz {\it et al.} (2005) reported 34 flares without coherent radio emission between 100 and 4000 MHz. The non-detections exclude ten cases apparently related to type I emission during strong radio noise storms. 

The 34 non-detections are here studied in the following way:
\begin{itemize}
\item The RHESSI data were analyzed for flare peculiarities.
\item The flare position was determined from RHESSI observations.
\item The RHESSI X-ray peak count rates in the bands 12 -- 25 keV to 25 -- 100 keV were determined.
\item GOES data were retrieved to determine more accurately the soft X-ray duration and peak.
\item The H$\alpha$ position reported in SGD was compared with X-ray position in critical cases.
\item The Phoenix-2 data were searched to higher sensitivity (five sfu) by integration.
\item Coincident type I emission at 164 MHz and 327 MHz were checked from Nan\c{c}ay Radioheliograph observations.
\item Reports in SGD from other radio observatories were studied for all events. In particular, we searched for radio bursts in the range 40 -- 100 MHz where Phoenix-2 did not observe.
\end{itemize}

The total bandwidth of Phoenix-2 (100 -- 4000 MHz) was searched more deeply for associated emission from the one-minute scale to subsecond resolution. The data allow easy detections of broadband emissions down to the lower limit at 100 MHz, but may be too sparse for narrowband events in meter waves, such as weak type I bursts. Type I bursts are only reported if they form drifting chains, a pattern that can easily be recognized in spectrograms and distinguished from terrestrial interference. They can last for days and thus are prone to spurious concidences. For this type we have complemented the data with the observations of the Nan\c{c}ay Radioheliograph published in Solar-Geophysical Data (SGD), providing both intensity and position at several frequencies. The basic observations were also complemented by observations of the NOAA GOES-8 satellite in soft X-rays (1.6 -- 12 keV and 3.1 -- 25 keV) using calibrated time profiles and H$\alpha$ observations from different sources listed in SGD to check the consistency of the flare position. 

\begin{table}
\caption{29 flares detected by RHESSI having no associated radio emission in the radio range 100 to 4000 MHz in the initial analysis of Phoenix-2 data. The peak time and radial position is determined from RHESSI 12 -- 25 keV observations. "Peak flux low" and "high" refer to the corrected 12 -- 25 keV and 25 -- 50 keV count rate, respectively. The background is subtracted. RHESSI data marked with * were measured in attenuator state 1, those with ** were measured with RHESSI at 12 -- 25 keV. "Radio type I" indicates that a noise storm was possibly associated. The columns "Radio 40 100" and "Radio 100 -- 4000" mark events that were associated with bursts between 40 -- 100 MHz (from SGD) and 100 -- 4000 MHz (using integrated Phoenix-2 data), respectively. }
\vskip2mm
\noindent
\begin{tabular}{ccccrrccccc}
\noalign{\hrule}
Nr.&Date&Peak&Radial&Peak&Peak&GOES&GOES&Radio&Radio&Radio \cr
&&time&position&flux&flux&class&duration&type I&40 -&100 -- \cr
&&&[$''$]&low&high&&[min]&&100&4000 \cr
\noalign{\hrule}
1&2002/03/29& 12:57:42 &364&220&0&C6.5&62&n&y&n \cr
2&2002/04/09&06:06:14&928&153&2&C7.9&38&y&n&n \cr
3&2002/04/30&08:20:44&956&690&60&M1.3&49&n&n&n \cr
4&2002/05/29&05:17:00&950&52&0&C8.1&155&n&n&n \cr
5&2002/05/30&05:20:24&952&251&4&M1.3&242&n&n&n \cr
6&2002/05/30&17:30:36&789&32&0&M1.6&113&y&y&n \cr
7&2002/07/05& 08:04:20&960&160&0&C7.8&45&n&n&n \cr
8&2002/07/11&14:18:08&839&462&23&M1.0 &24&n&n&n \cr
9&2002/08/11&11:44:24&970&537*&210*&C9.5&76&n&n&n \cr
10&2002/08/15&15:13:00&451&239&0&C6.9&17&n&y&n \cr
11&2002/08/23 & 06:24:56&934&1500&111&C8.6&33&n&y&n \cr
12&2002/08/23& 07:52:20&940&233&11&C6.1&10&n&y&n \cr
13&2002/08/23& 09:45:04&939&948&61&M1.5&197&n&n&n \cr
14&2002/08/23& 11:59:16&942&1172&54&M1.2&34&n&n&n \cr
15&2002/08/24&12:46:20&957&211&0&C5.4&50&n&y&n \cr
16&2002/08/29& 07:33:24&951&72&0&C7.2&83&n&n&n \cr
17&2002/09/06&16:26:18&960&1060&89&C9.2&14&n&n&n \cr
18&2002/11/08&12:22:38&476&308&0&C5.8&19&n&y&n \cr
19&2002/11/10&11:35:06&876&30&0&C5.1&13&n&n&y \cr
20&2002/11/21&11:02:10&379&131&0&C6.6&35&y&y&n \cr
21&2002/12/16& 09:27:50&680&342&0&C4.8&42&y&y&n \cr
22&2003/01/12& 14:24:34&952&47*&0&C6.1&100&n&n&n \cr
23&2003/02/01& 08:58:02&995&573*&34&M1.2&110&n&n&n \cr
24&2003/04/05&15:08:42&798&112&0&C4.9&165&y&y&n \cr
25&2003/04/24&15:47:06&958&78&6&C8.2&109&n&n&n \cr
26&2003/06/13& 06:37:38&949&561&9&M1.8&124&y&y&n \cr
27&2003/06/13& 07:59:46&981&5&0&C6.9&1**&y&n&n \cr
28&2003/06/13&09:30:42&936&15&0&C5.0&17&y&y&n \cr
29&2003/06/13&17:32:54&943&34&0&C4.6&9&y&n&n \cr
\noalign{\hrule}
\end{tabular}
\end{table}

\section{Results}
The survey of Benz {\it et al.} (2005) relied on an X-ray flare list produced by automatic routines without human intervention. Although the list has been checked manually, a second look revealed several misidentifications that resulted in radio non-detections. Four events turned out to be flawed enhancements due to an attenuator change, due to a terrestrial particle event, or due to two peaks in the same flare. One further event had to be eliminated because of associated radio emission in Phoenix-2 data overlooked previously. The remaining 29 flares are listed in Table I. They constitute a set of homogeneously selected events that is the basis of the following study. 

The duration of the GOES event from start to end was measured manually. This resulted in significantly different durations from the ones produced semi-automatically and reported in the GOES list. We have revised our earlier GOES classification to correspond to the official GOES flare list, which uses a longer integration time. Thus some of the events now have a class lower than the selection criterion of C5.0.  The soft X-ray flare position if reported by GOES 12 SXI and H$\alpha$ position (SGD) agrees with the RHESSI position. 

The radio information is presented in the last three columns in Table I. The first column lists the presence of type I radio bursts at the time of the flare at the X-ray position. The criterion for possible association was a radio noise storm reported by the Nan\c{c}ay Radioheliograph in SGD within half a solar radius at either 164 MHz or 327 MHz. Note that the association is questionable as noise storms are present on most days of high activity. A large radius for spatial coincidence is used as the sources of noise storms are generally much higher in the corona than coronal RHESSI sources. Thus an association would not necessarily mean a spatial coincidence. The SGD reports on bursts between 40 and 100 MHz are summarized in the next column. Only bursts coincident in time with the X-ray flare are reported. All of them are of type III. The final column presents the results of the reanalysis of the Phoenix-2 data at frequencies above 100 MHz using integrated data to enhance the sensitivity. Only one new radio burst of type III showed up (event number 19). Inspecting Ondrejov spectrometric observations in the 820 -- 2000 MHz and 2000 -- 4500 MHz range did not reveal any new radio bursts.

\section{Discussion}
Table I presents 29 events with little or no coherent radio emission. In 12 events of 29, radio emission was found below 100 MHz, thus originating at a plasma density below the plasma frequency of 100 MHz corresponding to an electron density below $1.3  \ 10^8$cm$^{-3}$. The observed radio bursts are consistent with plasma emission by electron beams escaping from the flare region (type III). 

The association with type I bursts is less conclusive. Obvious noise storm enhancements coinciding with the HXR event have been excluded already by Benz {\it et al.} (2005) from the list of radio-quiet flares. The nine events with possible noise storm association listed in Table I may contain spurious cases, in which the radio emission, occurring during a large fraction of the day, has little or no direct relation with the flare.

Even when the flares possibly associated with noise storms are excluded, there remain 13 flares with no coherent radio emission in the range from 40 MHz to 4000 MHz (only "n" entries in the last three columns of Table I). There is no doubt that flares with no coherent radio emission as observed by modern routine instrumentation do exist.

\subsection{Limb Events}
The most striking property of radio-quiet flares is their large radial position. According to Table I, 22 of 29 events (76\%) were found at a radius of more than 800$''$ from the center. The surface area visible from Earth beyond this radial position is only 16.6\%. Thus if the flares were distributed randomly, one would expect only 4.8 events beyond 800$''$. Assuming that the X-ray source is coronal and at a height of 0.1 solar radius, the expected number increases to 7.0. The preference for radio-quiet flares to occur near the limb (22 cases) is statistically highly significant. 

\subsection{Non-limb Events}
We have investigated the seven radio-quiet events at $R < 800''$ (events 1, 6, 10, 18, 20, 21, and 24 in Table I) in more detail. The comparison with the H$\alpha$ flare position did not reveal any inconsistencies. The $R<800''$ events have some remarkable properties in common: 

{\it (i)} All of the seven events show no hard X-ray flux above 25 keV. Thus they are weak and/or soft in HXR.  Out of 29 flares listed in Table I, 16 have this property. The chance to draw seven events with no $>$25 keV emission occurring in a random set is less than 1\%. 

{\it (ii)} All of the seven events have radio emission associated in the 40 -- 100 MHz range. Only 12 out of the 29 events have this property. Thus again, this characteristic is statistically significant with an error margin less than 1\%. 

{\it (iii)} Radio-quiet flares at $R<800''$ are relatively weak events having a GOES class $<$C7.0 with the exception of number 6 (2002/05/30; 17:30:36). This event at a marginal $R = 789''$ was reported as a gradual-rise-and-fall burst at 9500 MHz (Havana, SGD) and was also noticeable in the Phoenix-2 spectrum below 4000 MHz. Thus the radio emission must be classified as gyrosynchrotron emission and is therefore not reported in the last column of Table I.

{\it (iv)} Radio-quiet events at $R<800''$ have only one X-ray source (with one exception that however appears to be a loop rather than two footpoints). About half of the limb events show two or more HXR sources. 

Radio-quiet events at $R<800''$ do not statistically differ in soft X-ray duration from events at $R>800''$. Thus, they do not consist preferentially of gradual or extended flares. 

\section{Interpretations and Conclusions}
Solar flares lacking coherent radio emission have a pronounced preference to occur beyond $R>800''$. A likely interpretation is absorption or reflection of the coherent radio waves emitted close to the local plasma frequency. A further possibility is ducting in low-density troughs, guiding the radiation preferentially in vertical direction (Duncan, 1977). Waves propagating in nearly horizontal direction remain longer in a medium of high plasma frequency where they are more absorbed and more likely to be completely obscured by a region of $\omega > \omega_p$. The limb events at $R>800''$ amount to 76\% of all radio-quiet flares. These results thus suggest that a majority of the radio-quiet HXR flares reported in the literature have probably been limb flares. It is very likely that radio waves would have been observed from a position vertically above the source. 

Radio-quiet flares at $R<800''$, however, point to other causes for the lack of coherent radio emission. The investigated flares without emission above 100 MHz at $R<800''$ all had radio emission in the 40 -- 100 MHz range. We conclude that flares radio-quiet $>$100 MHz do not constitute evidence against propagating non-thermal electrons. 

Radio-quiet flares at $R<800''$ as seen from Earth are less likely to have resulted from radio wave absorption or reflection. They have several properties in common. Most remarkable is the soft HXR spectrum. Some of the non-detections at 25 keV may be due to the generally lower X-ray flux. In fact, radio-quiet flares at $R<800''$ are also weaker in GOES class. Thus they are generally less likely to have radio emission (as previously reported, see Section 1). 

Radio-quiet flares at $R<800''$ tend also to be softer in HXR emission. Most noticeable are events such as numbers 18 and 21 having a high count rate at 12 -- 25 keV but no HXR counts $>25$ keV. Thus we conclude that flares with softer HXR spectra are less likely to have coherent radio emission above 100 MHz. This may be explained by the prediction from the theory of velocity space instabilities ({\it e.g.} Benz, 2002) that a soft non-thermal electron distribution is less likely to become unstable by trapping or propagation. 

Assuming a random distribution, the total number of limb flares $R>800''$ is estimated to be 33.4 out of 201. Only 22 flares out of 201 have no radio emission and originate from the near the limb. Thus not all limb flares are absorbed or occulted in radio waves. In fact, radio events from near the limb are frequently reported, particularly for large GOES events. The absence of coherent radio emission in smaller limb flares should therefore not be taken as evidence for the complete lack of transparency of the corona for coherent radio waves from the flare energy release region. 

In conclusion, this study suggests that practically all flares with GOES class larger than C5 are associated with coherent radio emission if observed with current sensitivity from a position vertically above the source. Seen at the limb, the association rate with radio emission for flares smaller than GOES M2 class is significantly reduced. In addition to this limb effect, flares in a lower GOES class and having a soft HXR spectrum are less likely to be associated with coherent radio emission.

\begin{acknowledgements}
We thank the many people who have contributed to the successful operation of RHESSI and Phoenix-2 and to collecting this set of data, in particular C. Monstein and H. Meyer who have maintained and operated Phoenix-2 during the observing period. We have also made use of data from many other observatories, including the GOES satellite, the Nan\c{c}ay Radioheliograph, the radio spectrometers in Culgoora, IZMIRAN, Learmonth, Ondrejov, Potsdam, and San Vito, the Havana radiometer, as well as of other data listed in the Solar-Geophysical Data reports. Much of this work relied on the RHESSI Experimental Data Center (HEDC) supported by ETH Zurich (grant TH-W1/99-2). The construction of the Phoenix-2 spectrometer and the RHESSI work at ETH Zurich is partially supported by the Swiss National Science Foundation (grant 20-113556). R.B. would like to thank the Institute of Astronomy at ETH Zurich for financial support that enabled him to visit.
\end{acknowledgements}

{} 

\end{article}
\end{document}